\documentclass[onecollarge,natbib]{svjour2}
\bibpunct{[}{]}{;}{n}{}{,} % to get "[numbered]" references from natbib
\usepackage{graphicx}
\usepackage{epsfig}
\usepackage[usenames]{color}

\newcommand{\beq}{\begin{equation}}
\newcommand{\eeq}{\end{equation}}
\newcommand{\beqa}{\begin{eqnarray}}
\newcommand{\eeqa}{\end{eqnarray}}
\newcommand{\ba}{\begin{array}}
\newcommand{\ea}{\end{array}}

\journalname{Few Body Systems}

\begin{document}

\title{Dynamical properties of the unitary Fermi gas: \\
collective modes and shock waves}
\titlerunning{Dynamical properties of the unitary Fermi gas} 
\author{Luca Salasnich}
\institute{Luca Salasnich \at
Dipartimento di Fisica e Astronomia ``Galileo Galilei'' and CNISM,  
Universit\`a di Padova, Via Marzolo 8, 35131 Padova, Italy} 

\date{\today}

\maketitle

\begin{abstract}
We discuss the unitary Fermi gas made 
of dilute and ultracold atoms with an 
infinite s-wave inter-atomic scattering length. 
First we introduce an efficient Thomas-Fermi-von Weizsacker 
density functional which describes accurately various static 
properties of the unitary Fermi 
gas trapped by an external potential. Then, 
the sound velocity and the collective frequencies 
of oscillations in a harmonic trap are derived from 
extended superfluid hydrodynamic equations which are the Euler-Lagrange 
equations of a Thomas-Fermi-von Weizsacker action functional. 
Finally, we show that this amazing 
Fermi gas supports supersonic and subsonic shock waves.
\end{abstract}

\vskip 0.2cm 


\section{Introduction} 

For interacting fermions at very low temperature, 
far below the Fermi temperature $T_F$, the effects of quantum 
statistics become very important \cite{leggett}. 
When the densities of the two spin components are equal, and when the 
gas is dilute so that the range of the inter-atomic potential
is much smaller than the inter-particle distance, then the interaction 
effects are described by only one parameter: the s-wave scattering 
length $a_F$ \cite{leggett,giorgini}. The sign of $a_F$ determines the 
character of the gas. Fano-Feshbach 
resonances can be used to change the value and the sign of the 
scattering length, simply by tuning an external magnetic field. 
At resonance the scattering length $a_F$ diverges so that the gas 
displays a very peculiar character, being at the same 
time dilute and strongly interacting. In this regime all scales 
associated with interactions disappear from the problem 
and the energy of the system is expected to be proportional to that 
of a non interacting fermions system. This 
is called the unitary regime \cite{giorgini}. 

Recently it has been remarked \cite{bulgac1} that 
the unitary Fermi gas at zero temperature 
can be described by the density functional
theory. Indeed, different theoretical groups have proposed various 
density functionals. For example, Bulgac and Yu have introduced a superfluid 
density functional based on a Bogoliubov-de Gennes 
approach to superfluid fermions 
\cite{bulgac2,bulgac3}. Papenbrock and Bhattacharyya \cite{papenbrock} 
have instead proposed a Kohn-Sham density 
functional with an effective mass to take into account nonlocality effects. 
Here we adopt the extended Thomas-Fermi (ETF) functional of the 
unitary Fermi gas we have proposed few years ago \cite{salasnich,best} 
which is a functional of the fermions number 
density $n({\bf r})$ and of its gradient.
The total energy in the ETF functional 
contains a term proportional to the kinetic
energy of a uniform non interacting gas of fermions, plus 
a gradient correction of the von-Weizsacker 
form $\lambda \hbar^2/(8m) (\nabla n/n)^2$ \cite{von}. 
This approach has been adopted for studying the quantum hydrodynamics 
of electrons by
March and Tosi \cite{tosi}, and by Zaremba and Tso \cite{tso}.
In the context of the BCS-BEC crossover, the gradient term 
is quite standard \cite{v1,v2,v3,v4,v5,v6,v7,v8}. 
The main advantage of taking such a functional is the fact that, as it
depends only on a single function of the coordinates, there is no
limitation on the number of particles $N$ which it can treat. 
Other functionals, based on 
single-particle orbitals, require 
self-consistent calculations with a numerical load increasing with $N$. 

In this paper we review the last achievements obtained 
by using our ETF density functional and its time-dependent version 
\cite{salasnich,best}. Indeed we have successfully applied this 
density functional to investigate 
density profiles \cite{salasnich,best,salasnich-anci}, 
collective excitations \cite{salasnich-anci}, Josephson effect 
\cite{salasnich-anci2} and shock waves 
\cite{salasnich-shock} of the unitary Fermi gas. In addition, 
the collective modes of our EFT density functional have been used to study 
the low-temperature thermodynamics of the unitary Fermi gas 
(superfluid fraction, first sound and second sound) \cite{salasnich-thermo}
and also the viscosity-entropy ratio of the unitary Fermi gas 
from zero-temperature elementary excitations \cite{salasnich-ratio}. 

\section{BCS-BEC crossover and the unitarity limit} 

In 2002 the BCS-BEC crossover has been 
observed \cite{hara} with 
ultracold gases made of fermionic alkali-metal atoms. 
This crossover is obtained by changing with a Feshbach resonance 
the s-wave scattering length $a_F$ of the inter-atomic potential: 
if $a_F\to 0^-$ one gets the BCS regime of weakly-interacting Cooper pairs, 
if $a_F\to \pm \infty$ there is the 
unitarity limit of strongly-interacting Cooper pairs, and if 
$a_F\to 0^+$ one reaches the BEC regime of bosonic dimers. 
The many-body Hamiltonian of a two-spin-component Fermi system 
can be written as 
\beq 
{\hat H} = \sum_{i=1}^{N_{\uparrow}} 
\left( {{\hat p}_i^2\over 2m} + U({\bf r}_i) \right) 
+ \sum_{j=1}^{N_{\downarrow}} 
\left( {{\hat p}_j^2\over 2m} + U({\bf r}_j) \right) + 
\sum_{i,j} V({\bf r}_i-{\bf r}_j) \; ,    
\eeq
where $U({\bf r})$ is the external confining potential and 
$V({\bf r})$ is the inter-atomic potential. 
Here we consider $N_{\uparrow}=N_{\downarrow}$. 
The inter-atomic potential of a dilute gas can be modelled by 
a square well potential: 
\beq 
V(r) = \left\{ 
\ba{cc} 
-V_0 & \mbox{  for  } r < r_0 \\
0    & \mbox{  for  } r> r_0 \\
\ea
\right. 
\eeq
where $r_0$ is the effective radius. 
By varying the depth $V_0$ of the potential one changes 
the s-wave scattering length 
\beq 
a_F = r_0  
\left( 1 - {tan(r_0\sqrt{mV_0}/\hbar)\over 
r_0\sqrt{mV_0}/\hbar} 
\right) \; . 
\eeq
The crossover from a BCS superfluid ($a_F<0$) 
to a BEC of molecular pairs ($a_F>0$) 
has been investigated experimentally \cite{giorgini}, 
and it has been shown that 
the unitary Fermi gas ($|a_F|=\infty$) exists and is meta-stable. 
In few words, the unitarity regime of a dilute Fermi gas 
is characterized by 
\beq 
r_0 \ll n^{-1/3} \ll |a_F| \; .  
\eeq
Under these conditions the Fermi gas is called unitary Fermi gas. 
Ideally, the unitarity limit corresponds to 
\beq 
r_0 = 0 \quad\quad \mbox{and} \quad\quad a_F = \pm \infty \; . 
\eeq 
The detection of quantized vortices under rotation \cite{zw} 
has clarified that the unitary Fermi gas is superfluid. 

The only length characterizing the uniform unitary Fermi gas is 
the average distance between particles $d=n^{-1/3}$. 
In this case, from simple dimensional arguments, 
the ground-state energy per volume must be 
\beq 
{E_0\over V} = { \xi} 
{3\over 5} {\hbar^2 \over 2m} (3\pi^2)^{2/3} n^{5/3} = { \xi} \ 
{3\over 5} \ \epsilon_F \ n \; , 
\label{eos} 
\eeq 
with $\epsilon_F$ Fermi energy of the ideal gas, $n=N/V$ the total density,   
and $\xi$ a universal unknown parameter.  
Monte Carlo calculations and experimental data with dilute and 
ultracold atoms suggest \cite{giorgini} that ${ \xi}\simeq 0.4$. 

\section{Extended Thomas-Fermi density functional} 

The Thomas-Fermi (TF) energy functional \cite{giorgini} of 
the unitary Fermi gas in an external potential $U({\bf r})$ 
is given by 
\beq 
E = \int  d^3{\bf r} \ \Big[  { \xi} 
{3\over 5} {\hbar^2 \over 2m} (3\pi^2)^{2/3} n^{5/3}({\bf r}) 
+ U({\bf r}) \, n({\bf r}) \Big]  \ , 
\label{e-lda}
\eeq 
with $n({\bf r})=n_{\uparrow}({\bf r})+n_{\downarrow}({\bf r})$ 
total local density. The total number of fermions is 
\beq
N = \int d^3{\bf r} \ n({\bf r}) \; . 
\label{norma}
\eeq
By minimizing $E$ one finds 
\beq
{ \xi} {\hbar^2 \over 2m} (3\pi^2)^{2/3} n^{2/3}({\bf r}) 
 + U({\bf r}) = \bar{\mu} \; , 
\label{chem-lda}
\eeq 
with $\bar{\mu}$ chemical potential of the non uniform system. 
The TF functional can be extended to cure the 
pathological TF behavior at the surface, namely the fact that 
the density goes to zero at a finite distance from the center 
of the cloud. We add to the energy density the term 
\beq 
{ \lambda} {\hbar^2 \over 8 m} {(\nabla n)^2\over n}  
= { \lambda} {\hbar^2\over 2m} {(\nabla \sqrt{n})^2}  
\label{grad-term} \; .  
\eeq 
Historically, this term was introduced  by von Weizs\"acker \cite{von} 
to treat surface effects in nuclei. 
Here we consider ${ \lambda}$ as a {phenomenological parameter} 
accounting for the increase of kinetic energy due 
the spatial variation of the density. 
The new energy functional, that is the extended Thomas-Fermi (ETF) 
functional of the unitary Fermi gas, reads 
\beq 
E = \int d^3{\bf r} \ \left[ 
\lambda {\hbar^2 \over 8 m} {(\nabla n({\bf r}))^2\over 
n({\bf r})} + 
{ \xi} {3\over 5} {\hbar^2 \over 2m} (3\pi^2)^{5/3} n({\bf r})^{5/3}
+ U({\bf r}) \, n({\bf r}) \right] \  \; .  
\label{e-dft}
\eeq 
By minimizing the ETF energy functional one gets:
\beq
\left[ { \lambda} {\hbar^2\over 2m} \nabla^2 
+ { \xi} {\hbar^2 \over 2m} (3\pi^2)^{2/3} n({\bf r})^{2/3} 
+ U({\bf r}) \right] \sqrt{n({\bf r})} = 
\bar{\mu} \ \sqrt{n({\bf r})} \; .  
\label{chem-dft} 
\eeq 
This is a sort of stationary 3D nonlinear Schr\"odinger equation. 
In recent papers \cite{salasnich,best,salasnich-anci,salasnich-thermo,
salasnich-ratio} we have determined the parameters $\xi$ and $\lambda$
by fitting recent Monte Carlo results 
\cite{chang,blume,thermo-bulgac} for the energy of fermions confined in 
a spherical harmonic trap of frequency $\omega$ in this regime. 
The main conclusion is that the values 
\beq 
\xi = 0.40 \quad\quad \mbox{and} \quad\quad \lambda = 1/4 \;    
\eeq
fit quite well Monte Carlo data of the unitary Fermi gas.  
Having determined the parameters $\xi$ and $\lambda$,  
one can use our single-orbital density functional to calculate various 
properties of the trapped unitary Fermi gas. 
Ground-state energies and density profiles have been 
analyzed in Ref. \cite{salasnich,salasnich-anci}, showing a very good 
agreement with other theoretical aproaches \cite{blume,bulgac2}. 

\section{Generalized superfluid hydrodynamics}

Here we analyze the effect of the gradient term 
on the dynamics of the superfluid unitary Fermi gas. 
At zero temperature 
the low-energy collective dynamics of this fermionic gas 
can be described by the superfluid equations of 
inviscid hydrodynamics \cite{giorgini}:
\beqa 
{\partial n\over \partial t} + \nabla \cdot (n {\bf v}) = 0 \; , 
\label{hy1-s}
\\ 
m {\partial \over \partial t} {\bf v} + \nabla 
\Big[ { \xi}{\hbar^2\over 2m} (3\pi^2 n)^{2/3}
+ U({\bf r}) + {m\over 2} v^2 \Big] = 0 \; , 
\label{hy2-s}
\eeqa
where the velocity $\bf v$ is irrotational: 
$\nabla \wedge {\bf v}= 0$, i.e.  
\beq 
{\bf v}({\bf r},t) = {\hbar \over 2m} 
\nabla \theta({\bf r},t) \; , 
\eeq
with $\theta({\bf r},t)$ the phase of the condensate wave function 
of Cooper pairs. It is straightforward to show that 
these equations are the Euler-Lagrange 
equations of the following TF action functional 
\beq 
A = \int dt\, d^3{\bf r} \left[ 
{\hbar\over 2} {\dot{\theta}}\, n  + {\hbar^2\over 8m} 
(\nabla \theta)^2 \, n + 
{3\over 5} \xi {\hbar^2\over 2m} (3\pi^2)^{2/3} n^{5/3} 
+ U({\bf r}) \, n \right]   
\label{popov}
\eeq
Clearly, if the space-time variations of the phase are zero 
one recover the TF energy functional (\ref{e-lda}). 
From Eqs. (\ref{hy1-s}) and (\ref{hy2-s}) one finds \cite{giorgini}
the dispersion relation of low-energy collective modes of the
{uniform} ($U({\bf r})=0$) unitary Fermi gas in the form
\beq
\Omega_{col} = c_1 \ q  \; ,
\eeq
where $\Omega_{col}$ is the collective frequency, $q$ is the wave number and
\beq
c_1 =  \sqrt{{ \xi}\over 3} v_F
\eeq
is the first sound velocity, with $v_F = \sqrt{2\epsilon_F\over m}$ 
the Fermi velocity of a noninteracting Fermi gas.

The simplest extension of Eqs. (\ref{hy1-s}) and (\ref{hy2-s}) 
are the equations of extended \cite{tosi,tso} 
irrotational and inviscid hydrodynamics:
\beqa 
{\partial n\over \partial t} + \nabla \cdot (n {\bf v}) = 0 \; , 
\label{hy1}
\\ 
m {\partial \over \partial t} {\bf v} + \nabla 
\Big[ - \lambda {\hbar^2\over 2m} {\nabla^2 \sqrt{n}\over \sqrt{n}} + 
\xi{\hbar^2\over 2m} (3\pi^2 n)^{2/3}
+ U({\bf r}) + {m\over 2} v^2 \Big] = 0 \; .    
\label{hy2} 
\eeqa
These equations include the gradient correction and are the Euler-Lagrange 
equations of the following ETF action functional 
\beq 
A = \int dt\, d^3{\bf r} \left[ 
{\hbar\over 2} {\dot{\theta}} \, n 
+ {\hbar^2\over 8m} (\nabla \theta)^2 \, n + 
\lambda {\hbar^2 \over 8 m} {(\nabla n)^2\over n} + 
{3\over 5} \xi {\hbar^2\over 2m} (3\pi^2)^{2/3} n^{5/3} +  
U({\bf r}) \, n \right]  \; .  
\label{e-popov}
\eeq
By using Eqs. (\ref{hy1}) and (\ref{hy2}) one finds \cite{salasnich}
that the dispersion relation of low-energy collective modes of the
uniform unitary Fermi gas reads 
\beq
\Omega_{col} = c_1 \ q \ 
\sqrt {1+ {3{ \lambda} 
\over { \xi}} 
\big( {\hbar q \over { 2 m v_F}}\big)^2 } \; ,   
\label{sound}
\eeq 
where the cubic correction depends on the ratio ${ \lambda}/{ \xi}$. 

In the case of spherically-symmetric harmonic confinement  
\beq
U({\bf r})= {1\over 2} m \omega^2 r^2 
\eeq
we study numerically the collective modes of the unitary Fermi gas 
by increasing the number $N$ of atoms by means of 
Eqs. (\ref{hy1}) and (\ref{hy2}). 
As predicted by Y. Castin \cite{castin}, 
the frequency $\Omega_0$ of the monopole mode (breathing mode)   
and the frequency $\Omega_1$ dipole mode (center of mass 
oscillation) do not depend on $N$: 
\beq 
\Omega_0 = 2 \omega  \quad\quad \mbox{ and } \quad\quad \Omega_1 
= \omega \; .   
\eeq
We find instead that the frequency $\Omega_2$ of the quadrupole ($l=2$) mode 
depends on $N$ and on the choice of the gradient 
coefficient $\lambda$. We solve numerically \cite{salasnich-anci}
Eqs. (\ref{hy1}) and (\ref{hy2}) with the initial condition 
\beq 
n({\bf r},t=0) = n_{gs}({\bf r}) \, e^{i\epsilon (2z^2 - x^2 - y^2)}  
\eeq
to excite the quadrupole mode, where $n_{gs}({\bf r})$ is the ground-state 
density profile and $\epsilon$ a small parameter, 

\begin{figure}
\begin{center}
\epsfig{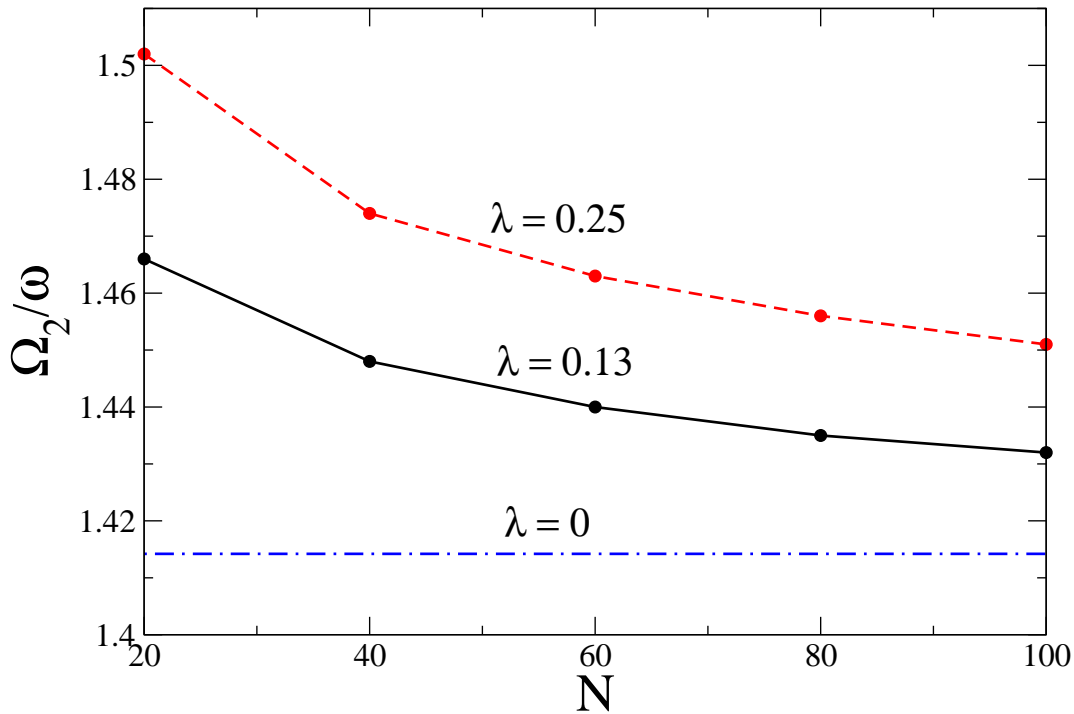}
\end{center}
\caption{Quadrupole frequency $\Omega_2$ of 
the unitary Fermi gas with $N$ atoms under harmonic 
confinement of frequency $\omega$. Three different 
values of the gradient coefficient ${ \lambda}$. 
For $\lambda=0$ (TF limit): $\Omega_2=\sqrt{2} \omega$. 
Figure adapted from Ref. \cite{salasnich-anci}.}
\label{fig1}
\end{figure}

The results are shown in Fig. \ref{fig1} where we plot 
the quadrupole frequency $\Omega_2$ as a function of the number 
$N$ of atoms for three values of the gradient coefficient $\lambda$. 
The trend shown in the figure is captured by the formula 
\beq 
\Omega_2 = \omega \ \sqrt{2 + 6\, \alpha\, {\lambda \over N^{2/3}} 
\over 1 + {3\over 2} \alpha\, {\lambda \over N^{2/3}}} \; , 
\eeq
where $\alpha=3(6/5)^{3/2}(3\pi^2)^{2/3}\xi/5$. This expression 
is easily derived by using a time-dependent 
Gaussian variational approach \cite{sala-tdva} in the ETF action functional 
(\ref{e-popov}). In the limit $N\to \infty$ it gives the TF result 
$\Omega = \sqrt{2}\omega$ \cite{giorgini}, 
while in the limit $N\to 0$ it gives 
$\Omega = 2 \omega$, which is the quadrupole oscillation frequency 
of non-interacting atoms \cite{giorgini}. 

\section{Shock waves} 

One of the basic problems in physics is how density perturbations 
propagate through a material. In addition to the well-known 
sound waves, there are shock waves characterized by 
an abrupt change in the density of the medium: they produce,  
after a transient time, an extremely large density gradient (the shock).
Shock waves are ubiquitous and have been studied in many different physical 
systems \cite{landau,whitham}. Here we investigate the formation 
and dynamics of shock waves in the unitary 
Fermi gas by using the zero-temperature equations of 
generalized superfluid hydrodynamics, inspired by the  
very recent observation of nonlinear hydrodynamic waves 
in the collision between two strongly 
interacting Fermi gas clouds of $^6$Li atoms \cite{thomas}. 

Let us consider the unitary Fermi gas 
with constant density $\bar{n}$ with $U({\bf r})=0$. 
Experimentally this configuration can be obtained 
with a very large square-well potential (or a similar external trapping), 
such that in the model one can effectively impose 
periodic boundary conditions instead of the vanishing ones. 
A density variation along the $z$ axis with respect to the uniform 
configuration $\bar{n}$ can be experimentally created 
by using a blue-detuned (bright perturbation) 
or a red-detuned (dark perturbation) laser beam \cite{leggett,giorgini}. 
In practice, we perform the following factorization 
\beq 
n({\bf r},t)= n_{\bot}(x,y) \, n_{\parallel}(z,t) \; , 
\eeq
by imposing also 
\beqa 
n_{\bot}(x,y) = \bar{n}_{\bot}
\\
n_{\parallel}(z,t) = \bar{n}_{\parallel} \; \rho(z,t) 
\eeqa
such that  
\beq 
n({\bf r},t) = \bar{n} \, \rho(z,t) 
\label{ansatz}
\eeq
where $\bar{n}=\bar{n}_{\bot} \bar{n}_{\parallel}$, and 
$\rho(z,t)$ is the relative density, i.e. the 
localized axial modification with respect to the uniform 
density $\bar{n}$. We impose periodic boundary 
conditions along the $z$ axis, namely $\rho(z=L_z,t)=\rho(z=-L_z,t)$, 
with $2L_z$ the axial-domain length. 
We set ${\bf v}({\bf r},t)=(0,0,v(z,t))$ with $v(z,t)$ 
the velocity field {such that $v(z=L_z,t)=v(z=-L_z,t)$.} 
Moreover we impose that the initial localized 
wave packet satisfies the boundary 
conditions $\rho(z=\pm L_z,t=0)=1$ and $v(z=\pm L_z,t=0)=0$. 
Because the dimensional reduction is done 
assuming the uniformess in $x$,$y$ directions, we shall 
consider the propagation of a plane wave along the $z$ axis. 

\begin{figure}
\begin{center}
\epsfig{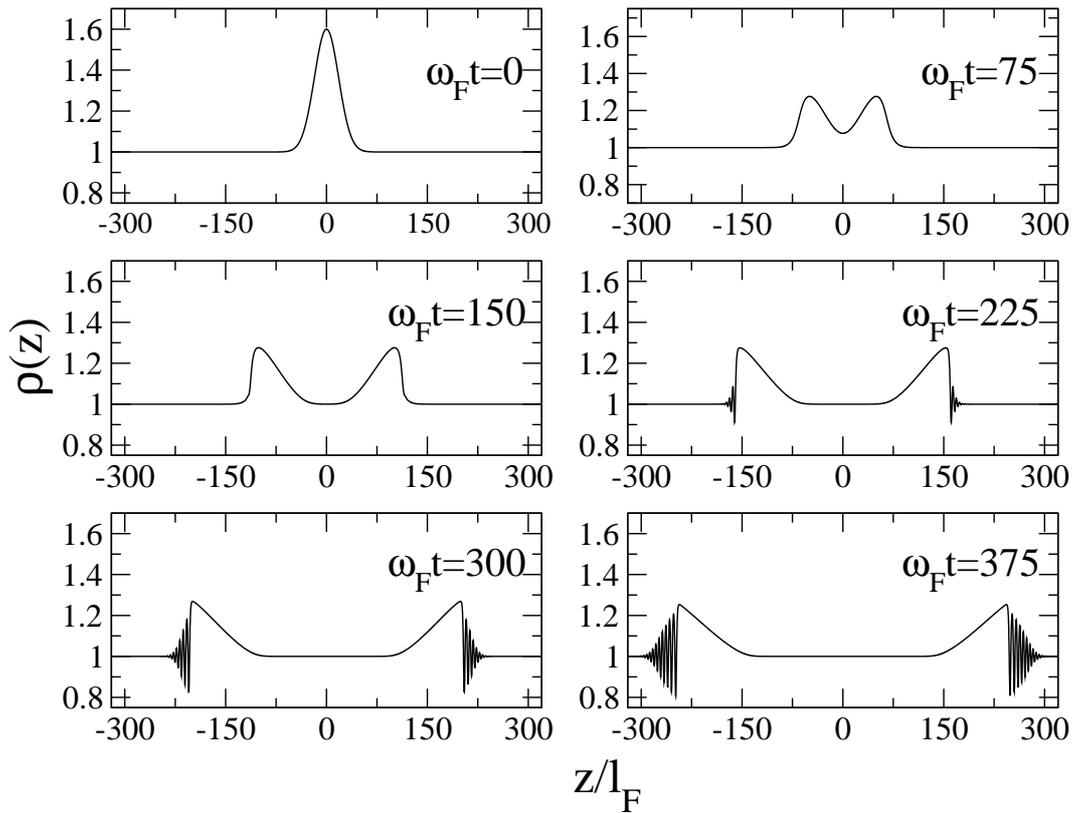}
\end{center}
\caption{Time evolution of supersonic shock waves. 
Initial condition with $\sigma/l_F=18$ and $\eta=0.3$. 
The curves give the relative density profile $\rho(z)$ 
at subsequent frames, where 
$l_F=\sqrt{\hbar^2/(m\epsilon_F)}$ is the Fermi length 
and $\omega_F=\epsilon_F/\hbar$ is the Fermi frequency. 
Figure adapted from Ref. \cite{salasnich-shock}.}
\label{fig2}
\end{figure}

Inserting Eq. (\ref{ansatz}) into Eqs. (\ref{hy1}) and (\ref{hy2}) 
one finds the 1D hydrodynamic equations for 
the axial dynamics of the superfluid, given by  
\beqa 
{\dot \rho} + v \rho' + v' \rho = 0 \; , 
\label{uf1}
\\
{\dot v} + v v' +  {c_{ls}(\rho)^2\over \rho} \rho' = 0 \; ,    
\label{uf2}
\eeqa 
where dots denote time derivatives, primes space derivatives,  
and 
\beq 
c_{ls}(\rho) = c_s \rho^{1/3} 
\label{nice-cs}
\eeq
is the local sound velocity, with 
$c_s=c_{ls}(1)=\sqrt{\xi/3}v_F$ the bulk sound velocity, 
$v_F=\sqrt{2\epsilon_F/m}$ is bulk Fermi velocity 
and $\epsilon_F={\hbar^2\over 2m }(3\pi^2 \bar{n})^{2/3}$ 
the bulk Fermi energy.

\begin{figure}
\begin{center}
\epsfig{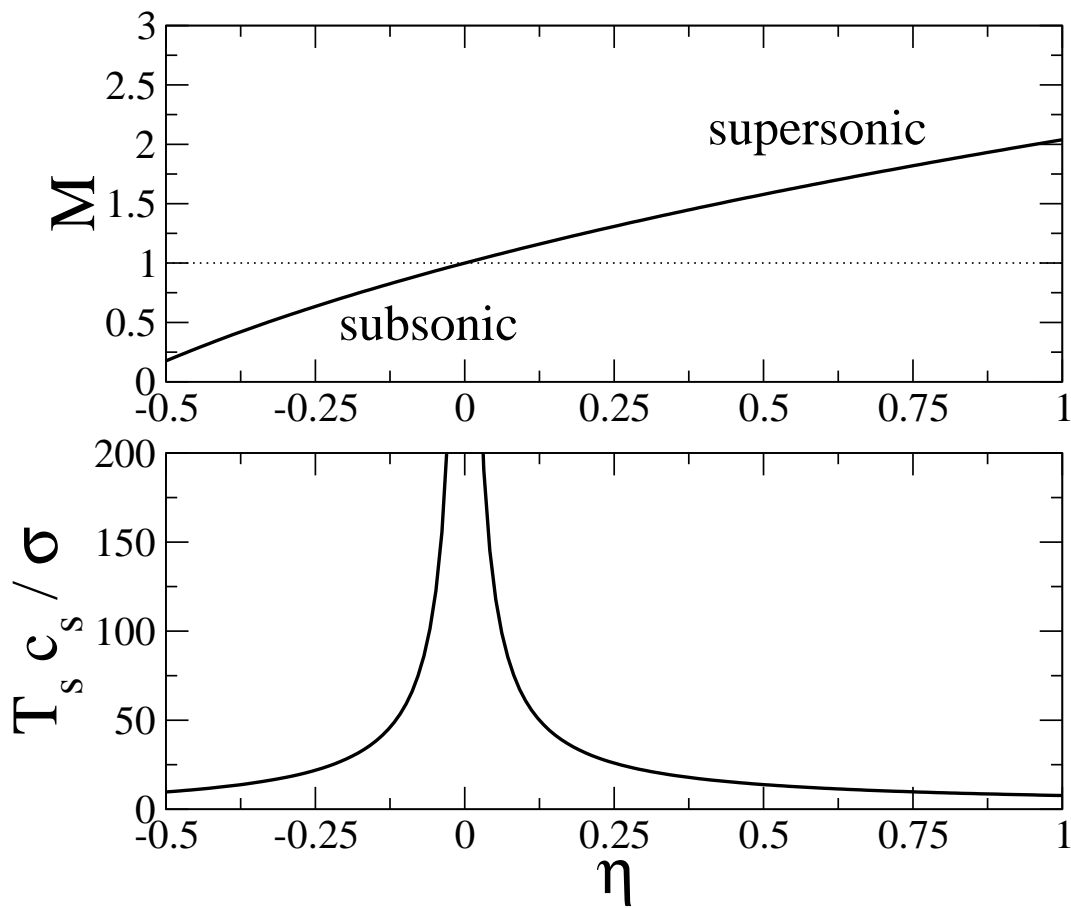}
\end{center}
\caption{Upper panel: Mach number $M=v_{max}/c_s$ as a function of the 
amplitude $\eta$ of the perturbation (solid line). 
Lower panel: period $T_s$ of formation (breaking time) 
of the shock-wave front 
as a function of the amplitude $\eta$ of the perturbation. 
$T_s$ is in units of $\sigma/c_s$, where $\sigma$ is the width 
of the perturbation and $c_s=\sqrt{\xi/3}v_F$ 
is the bulk speed of sound, with $v_F$ the Fermi velocity. 
Figure adapted from Ref. \cite{salasnich-shock}.}
\label{fig3} 
\end{figure}

We solve Eqs. (\ref{uf1}) and (\ref{uf2}) by using a Crank-Nicolson 
finite-difference predictor-corrector algorithm \cite{sala-numerics} 
with the initial condition given by 
\beq 
\rho(z,t=0) = 1 + 2\eta \ e^{-z^2/(2\sigma^2)}  \; , 
\label{initial-rho} 
\eeq
and $v(z,t=0)=0$, where $\bar{n}$ is the bulk density. 
In Fig. \ref{fig2} we plot the time evolution 
obtained with $\sigma/l_F=18$ and $\eta=0.3$, with 
$\sqrt{\hbar^2/(m\epsilon_F)}$ the Fermi length of the bulk system. 
The figure displays the density profile $\rho(z)$ 
at subsequent times. Note the splitting 
on the initial bright wave packet into two bright travelling waves moving 
in opposite directions. 
There is a deformation of the two waves with the formation 
of a quasi-horizontal shock-wave front. Eventually, 
this front spreads into wave ripples. We have carefully checked 
that these ripples are not an artefact of the numerical scheme. 
Notice that before the shock 
both amplitude and velocity of the two maxima 
of the two waves are practically constant during time evolution. 
In particular, as we have recently shown \cite{salasnich-shock} 
that the amplitude of the extrema can be written  
$A(\eta) = 1+\eta$ while the velocity of the extrema reads 
\beq 
V(\eta) = \pm c_s \left(4(1+\eta)^{1/3}-3 \right)\; . 
\label{nice-vmax}
\eeq
Taking $\eta =0$ the velocity of the impulse 
extrema reduces to the sound velocity: $V(0)=c_s=\sqrt{\xi/3}v_F$. 
Moreover, bright perturbations ($\eta >0$) move 
faster than dark ones ($\eta < 0$), and the Mach number 
$M=V(\eta)/V(0)$ of these perturbations 
in the unitary Fermi gas is simply 
\beq 
M = 4(1+\eta)^{1/3}-3 \; .   
\eeq 
For $M>1$, which means $\eta>0$ (bright perturbation),  
one has supersonic waves, 
while for $0\leq M<1$, which means $\eta<0$ (dark perturbation),  
one has subsonic waves.  In the upper panel 
of Fig. \ref{fig3} we plot the Mach number $M$ as function of the 
amplitude $\eta$ of the perturbation. Note that since $2\eta$ 
is the amplitude of the initial condition, 
see Eq. (\ref{initial-rho}), the region $\eta \leq -1/2$ is unphysical. 

Let us consider a bright perturbation ($\eta>0$) moving to the right. 
The speed of impulse maximum $V(\eta)$ 
is bigger than the speed of its tails $V(0)$. 
As a result the impulse self-steepens in the direction 
of propagation and a shock wave front 
takes place. The breaking-time $T_s$ required for such a process can 
be estimated as follows: the shock wave front appears 
when the distance difference traveled by lower and upper impulse 
parts is equal to the impulse half-width $\sigma/2$. 
This criterion gives \cite{salasnich-shock}  
\beq 
T_s = {\sigma \over 2c_s ((1+\eta)^{1/3}-1) } 
\; .  
\label{nice-ts}
\eeq 
In the lower panel of Fig. \ref{fig3} we plot the period $T_s$ as a function 
of the amplitude $\eta$ of the perturbation. The figure shows that 
as $\eta$ goes to zero the period $T_s$ goes to infinity; 
in fact, in this limit the shock wave reduces to a 
sonic wave (sound wave) which does not produce a shock. 

\section{Conclusions} 

We have shown that the ETF functional of the unitary Fermi gas 
can be used to determine ground-state properties of the 
system in a generic external potential $U({\bf r})$. Moreover, 
the time-dependent version of this EFT function, that is 
the generalized hydrodynamics equations, can be applied 
to calculate sound waves, collective modes and also shock waves. 
Our generalized superfluid hydrodynamics is reliable when 
the effects of the temperature $T$ are negligible, namely 
if $T\ll T_c$, with $T_c$ the 
critical temperature of the superfluid-normal phase transition 
($T_c\simeq 0.2\ T_F$, and $T_F \simeq 10^{-7}$ Kelvin 
for dilute alkali-metal atoms). As previously discussed, 
recently an observation of nonlinear hydrodynamic waves has 
been reported in collisions between two strongly interacting Fermi gas 
clouds of $^6$Li atoms \cite{thomas}. 
The experiment shows the formation of density gradients, which are
nicely reproduced by hydrodynamic equations with a phenomenological 
viscous term \cite{thomas}. Nevertheless, the role of dissipation is 
questionable in this case since the ultracold unitary
Fermi gas is known to be an example of an almost perfect 
fluid \cite{cao}. We plan to simulate the collision 
between two strongly interacting 
Fermi-gas clouds of $^6$Li atoms by using our single-orbital
time-dependent density functional. In particular, 
we aim to demonstrate that it is possible to reproduce 
the recent experimental results 
without the inclusion of a phenomenological dissipative term. 
Another interesting related problem 
is the determination of the surface tension of the unitary 
Fermi superfluid. We shall consider a semi-infinite
domain, derive the density profile of the system by using 
our density functional, and obtain the surface tension as the 
grand potential energy difference between the
actual configuration and the uniform asymptotic one. 
Finally, we shall compare our result with
previous determinations based on microscopic calculations 
of the normal-superfluid interface in
population-imbalanced Fermi gases \cite{bojan}.

\begin{acknowledgements}
The author thanks Sadhan Adhikari, 
Francesco Ancilotto, Nicola Manini, and Flavio Toigo 
for useful discussions and suggestions. 
\end{acknowledgements}

\end{document}